\documentclass{nws}

\expandafter\let\csname equation*\endcsname\relax
\expandafter\let\csname endequation*\endcsname\relax

\usepackage{amsmath}
\usepackage{natbib}
\usepackage{mathrsfs}
\usepackage{tikz}
\usepackage{amssymb}

\newcommand{\DDg}[4]{{#1}_{#2}^{#3}(#4)} 
\newcommand{\DD}[3]{{#1}_{#2}^{#3}} 

\title[Dynamic Network Prediction]
{Dynamic Network Prediction}

\author[R. Goyal]
{Ravi Goyal\\
Mathematica\\
\and \\
Victor De Gruttola\\
Department of Biostatistics, Harvard School of Public Health 
\email{rgoyal@mathematica-mpr.com}}

\begin{document}

\label{firstpage}

\maketitle

\begin{abstract}
We present a statistical framework for generating predicted dynamic networks based on the observed evolution of social relationships in a population. The framework includes a novel and flexible procedure to sample dynamic networks given a probability distribution on evolving network properties; it permits the use of a broad class of approaches to model trends, seasonal variability, uncertainty, and changes in population composition. Current methods do not account for the variability in the observed historical networks when predicting the network structure; the proposed method provides a principled approach to incorporate uncertainty in prediction. This advance aids in the designing of network-based interventions, as development of such interventions often requires prediction of the network structure in the presence and absence of the intervention. Two simulation studies are conducted to demonstrate the usefulness of generating predicted networks when designing network-based interventions.  The framework is also illustrated by investigating results of potential interventions on bill passage rates using a dynamic network that represents the sponsor/co-sponsor relationships among senators derived from bills introduced in the US Senate from 2003-2016. 

\end{abstract}

\section{Introduction}

Complex social systems in which individual-level outcomes of interest are interdependent are increasingly represented as networks. In some systems, such as those involving transmission of sexual diseases, the dependencies among people--represented as nodes in a network--are not permanent, but form and dissolve over time, leading to changes in the network topology. In this report, we present an approach to predict the topological evolution of the network based on observed historical network data. Specifically, we assume the evolving network is observed at discrete times $t={0,\cdots,k}$ and we want to predict the network for times $t=\{k+1,k+m\}$. The primary advance in our proposed approach is a method to generate a dynamic network, i.e., networks for times $t=\{k+1,k+m\}$, from a broad class of probability distributions. The method allows investigators to model long-term and seasonal trends in the evolution of the network structure--as observed in networks at times $t={0,\cdots,k}$--and to use these models to generate predicted networks in ways that incorporate uncertainty in the topology of the predicted networks.  

There are several frameworks for modeling the dynamic network that enable generating predicted networks. One such framework consists of the stochastic actor-oriented models, models that define continuous time Markov processes to govern the formation or dissolution of edges \citep{snijders2017stochastic, stadtfeld2018micro}. Another common framework embeds network information into a low dimensional latent space \citep{sewell2015latent}. A third models dynamic networks through extensions to the static exponential random graph model (ERGM) \citep{FS86}. These include temporal ERGMs (TERGMs) and separable TERGMs (STERGMs) \citep{hanneke2010discrete, krivitsky2013separable}. There is also research based on matrix completion for estimating the network structure of partially observed static networks that incorporates uncertainty; we are, however, not aware of extensions to dynamic networks \citep{chatterjee2015matrix}. To our knowledge, current methods do not allow estimates of the level of uncertainty in the predicted network structure for times $t=\{k+1,k+m\}$ to be based on the variability in the observed networks at times $t={0,\cdots,k}$. Incorporating the observed historical variability of network properties is important for predicting network structure.

Significant methodological challenges exist in designing interventions that modify network structure.  A primary challenge is the lack of general theory that connects network properties to outcomes of network processes, such as disease propagation.   Considerable research has been devoted to investigating this relationship; however, the focus has mostly been on static networks. \citet{pellis2015eight} commented on the need for additional research on dynamic networks in the context of epidemiological investigations. In the absence of theory, modeling time trends in outcomes requires modeling the entire evolution of the network. Therefore, assessing the potential impact of an intervention requires prediction of the network structure in the presence and absence of the intervention. Such an  approach was used in the design and monitoring of a large randomized community controlled trial, the Botswana Combination Prevention Program (BCPP), which investigates whether implementation of a combination of prevention interventions reduces HIV incidence \citep{Wang13}. 

Our proposed approach allows investigators to predict the dynamic network structure using historical data on the network prior to the implementation of the intervention. It also enables investigators to adjust the probability distribution of network properties targeted for modification by  the intervention in order to generate predicted networks in its presence. Comparing  results from simulations modeling processes operating on networks in the presence and absence of an intervention  allows for the evaluation  of its potential impact.

Our proposed method is described in sections 2 to 4; comparison with (S)TERGMs in section 5; and applied to simulation studies and an analysis of an observed network in sections 6 and 7, respectively.  Specifically, section 2 introduces network terminology, section 3 provides a conceptual framework for generating predicted dynamic networks, and section 4 provides details of the method to sample dynamic networks. Section 5 provides technical details of (S)TERGMs and a simulation study comparing the proposed method with STERGMs. Section 6 demonstrates how generating predicted dynamic networks is useful in the investigation of interventions designed to modify network properties.  Section 7 illustrates the method through an analysis of a dynamic network that represents the sponsor/co-sponsor relationships among senators indicated from bills introduced in the US Senate from 2003-2016. The network structures for the $108^{th}-113^{th}$ Senates (2003-2014) are used to predict the dynamic network structure for the $114^{th}$ Senate (2015-2016).  Section 7 also demonstrates the usefulness  of the method in predicting hypothetical effects of  interventions that modify the level of bipartisan support of bills--specifically  the  number of  ties between senators of different parties in the sponsor/co-sponsor networks. Section 8 discusses the limitations of the procedure and provides suggestions for further research. An R library to use the proposed methods is available by request.\footnote{The currently available R library CCMnet on CRAN will be updated to include the presented methods.} 

\section{Network Terminology}\label{Terminology}

We represent a population and connections among its members at time $t$ as a network, denoted as $g_t = (V_t, E_t)$, where the sets $V_t$ and $E_t$ represent individuals and their connections at time $t$. The network $g_t$ can be equivalently represented as a binary adjacency matrix with dimensions equal to the size of the set $V_t$; therefore, $g_t$ has dimensions $\vert V_t \vert \times \vert V_t \vert$, where $\vert Z \vert$ denotes the size of set $Z$. Let $g_t[i,j] = 1$ indicate that there is a relationship between individuals $i \in V_t$ and $j \in V_t$ at time $t$, i.e., $(i,j) \in E_t$, while $g_t[i,j] = 0$ indicates that there is no relationship, i.e., $(i,j) \notin E_t$. Let $n_t(i)$ be the individuals with connections to individual $i$, i.e., $n_t(i) = \{j : g_t[i,j] = 1\}$. Let $\mathcal{G}_t$ be the entire space of networks with $V_t$ as nodes.

Mixing patterns describe the tendency for individuals in networks to be connected to others that are like (or unlike) them based on their particular characteristics;  we consider only discrete characteristics. Let $\DDg{m}{i}{}{g_t}$ represent the vector of discrete characteristics for individual $i$ in network $g_t$. We consider only a single characteristic, political party affiliation; but the methods and formulas permit investigation of multiple characteristics. Let $\DDg{m}{}{}{g_t} = (\DDg{m}{1}{}{g_t},\cdots, \DDg{m}{\vert V_t \vert}{}{g_t})$ be a vector containing the characteristics of all individuals. The characteristic distribution, denoted as $\DDg{M}{}{}{g_t}$, is a vector representing the number of individuals with these characteristics over all individuals; the $k^{th}$ entry represents the number of individuals having characteristic $k$, i.e., $\DDg{M}{k}{}{g_t} = \sum_{i = 1}^{\vert V_t \vert} I_{\{\DDg{m}{i}{}{g_t} = k\}}$. 

We define a discrete-time dynamic network as a sequence of 2 or more static networks representing the evolution of social relationships in a population; we will refer to a discrete-time dynamic network simply as a dynamic network in this report. We refer to the number of static networks in a dynamic network as its length. We consider a dynamic network model in which the network at time $t$ is a single draw from a probability distribution that conditions on the networks at times $t-1,\cdots,t-k$, denoted as $P_{\mathcal{G}_t}(G_t = g_t \vert g_{t-1},\cdots, g_{t-k})$ where $G_t$ is a random variable with support $\mathcal{G}_t$. The probability distribution is based on network properties that characterize salient features of the evolving network. We denote a collection of network properties of a system as  essential  if the collection cannot be reduced and still adequately characterize the system.  Although there have been recent methodological advances to assess  whether a network model adequately characterizes a system \citep{ hunter2008goodness, hanneke2010discrete, schweinberger2012statistical}, additional research in this area is still needed; in practice, the assessment may require simulation studies and guidance from subject matter experts.

Define $\eta(g_t \vert g_{t-1} \cdots, g_{t-k})$ to be the function that maps $g_t \in \mathcal{G}_t$ to the values of the essential network properties conditional on $g_{t-1},\cdots, g_{t-k}$. Let $c_{x} = \eta^{-1}(x \vert g_{t-1},\cdots, g_{t-k})$, the inverse image of the function of $\eta(g_t \vert g_{t-1},\cdots, g_{t-k})$, i.e., $c_x = \{g_t : \eta(g_t \vert g_{t-1},\cdots, g_{t-k}) = x, g_t \in \mathcal{G}_t\}$; we refer to $c_x$ as a congruence class of $\mathcal{G}_t$ for the specified essential network properties. Let $P_{\mathcal{C}_t}(C_t)$ be the probability distribution of essential network property values where $C_t$ is a random variable for the vector of real values that are associated with the congruence classes of $\mathcal{G}_t$. The relationship between $P_{\mathcal{G}_t}$ and $P_{\mathcal{C}_t}$ is shown below:

\begin{equation} \label{eq:networkprob_static_pc}
P_{\mathcal{C}_t}(\eta(g_t \vert g_{t-1} \cdots, g_{t-k})) = \sum_{g \in c_{\eta(g_t \vert g_{t-1} \cdots, g_{t-k})}} P_{\mathcal{G}_t}(g \vert g_{t-1},\cdots, g_{t-k}).
\end{equation}

\section{Dynamic Network Prediction Framework}\label{ConceptualFramework}

The proposed network prediction method has three components. The first  identifies essential network properties by defining the mapping $\eta: \mathcal{G}_t \rightarrow \mathbb{R}^p$, where $p$ is the number of essential network properties.  As discussed in \citet{krivitsky2013separable}, it may be useful to  specify two types of essential network properties; one of these aids in characterizing the cross-sectional properties of the network, and the other, its longitudinal properties.  We follow this approach and refer to cross-sectional and longitudinal properties as static and dynamic essential network properties, respectively.    The former consists of properties that are calculable based only on $g_t$; the latter are those that require information on previous networks in order to be calculated. Let $\eta_s(g_t)$ and $\eta_d(g_t \vert \cdots, g_{t-k})$ denote the mapping from $\mathcal{G}_{t}$ to the values of the static and dynamic essential network properties, respectively, where $k$ denotes the number of previous networks necessary to compute the dynamic essential network properties. Note that $\eta(g_t \vert g_{t-1} \cdots, g_{t-k}) = (\eta_s(g_t), \eta_d(g_t \vert g_{t-1} \cdots, g_{t-k}))$.  

The distinction between the two types of essential network properties can be illustrated using the US sponsor/co-sponsor network.  Static properties provide information on the number of relationships between US Senators at each time point, for example during January 2016; by contrast, dynamic properties  provide information on the number of relationships in January 2016 that persisted to February 2016, i.e., the rate of evolution in the dynamic network.  Dynamic properties can capture the rate of evolution in the system--not only overall, but also with regard to specific types of relationships (e.g., relationships between members of the same political party or  between members of different parties). 

There are both practical and theoretical reasons for stratifying the essential network properties into two categories. As mentioned by \citet{krivitsky2013separable}, the practical reason is that information about the static and dynamic properties of a network are often derived from distinct sources. Therefore, it may be necessary to model the functional form of these properties separately. Regarding the theoretical reason, the functional form of the equation necessary to sample from a DCCM (shown later in Equation ~\ref{eq:prob_accept_DCCM}) relies on the model for the specific essential network properties; the approaches to derive or estimate the functional forms differ between static and dynamic essential network properties (details are provided in Section 7 and Appendix).

Recent statistical advances in dynamic networks complement our proposed method as they can be used to guide selection of the essential network properties \citep{hanneke2007discrete, hanneke2010discrete, krivitsky2013separable, paul2013hierarchical, Snijders96}. Also of importance is  recent work on assessing goodness-of-fit (GOF) for static networks \citep{hunter2008goodness} and for temporal networks \citep{hanneke2010discrete, schweinberger2012statistical}. These advances can aid in identifying and validating the selected set of essential network properties; as mentioned above, additional research is needed in the area of assessing GOF. We use the method as proposed in \citet{hanneke2010discrete} to assess GOF for the US Senate Bill data. 

The second component is modeling and predicting essential network properties, i.e., specifying $P_{\mathcal{C}_t}$. The framework presented in this paper provides the flexibility needed to specify the probability of observing a network with particular values for the selected essential network properties using a range of techniques, including techniques for modeling evolving trends and seasonal variability.

The first component (identifying essential network properties) does not provide a probability of observing a network at time $t$, but only specification of the properties that are used to compute that probability. Therefore, our framework requires two distinct decisions about the use of prior networks for modeling dynamic networks. The first is the number of prior networks that are used to define the dynamic essential network properties. Use of only the previous network would require that the persistence of an edge $(i,j) \in g_{t-1}$ until time $t$ does not depend on whether the edge was present at time $t-2$ (or any other earlier time). The second  is the collection of observed historical networks to use to estimate the joint distribution of essential network properties at time $t$, i.e., estimate $P_{\mathcal{C}_t}$. The choice of which prior networks to use for each of these two decisions can differ. For example, we might make the assumption stated above--that the persistence of a relationship at time $t$ depends only its presence at time $t-1$; nonetheless, the estimate for the number of relationships at time $t$ and of the subset of relationships that existed at both times $t$ and $t-1$ might usefully be based on historical averages dating back years or decades.

\begin{figure}
\centerline{\includegraphics[scale=0.5]{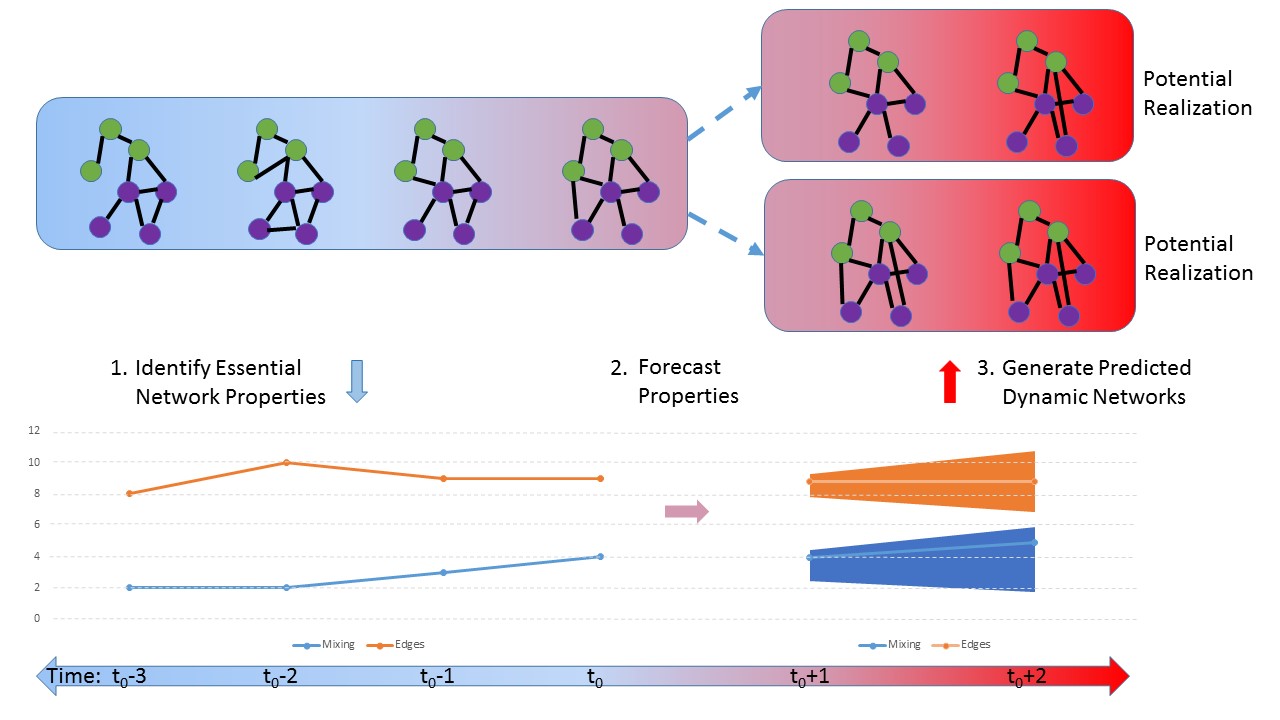}}
\caption{A conceptual illustration of the prediction framework consisting of three components: identify essential network properties, forecast properties, and generate predicted networks.}
\label{figCF}
\end{figure}

The third component is generation of networks according to  the probability distribution $P_{\mathcal{G}_t}$, which is based on the predicted distribution of the essential network properties, $P_{\mathcal{C}_t}$; this relationship is shown in equation (\ref{eq:networkprob_static_pc}). In essence, this component maps back to the network space $\mathcal{G}_t$ from $\mathbb{R}^p$ by sampling networks from the probability distribution $P_{\mathcal{G}_t}$. The three components of this framework are illustrated in figure~\ref{figCF}.

\section{Dynamic Congruent Class Model}

\subsection{Congruence Class Model}

To maximize the flexibility of the methods used to estimate the predictive distribution for the network properties, we propose a general procedure to generate networks based on a model by \citet{GBD13}; we refer to it as the Congruence Class Model (CCM). We extend the CCM, a method for static network generation, to dynamic networks and refer to this extension as the Dynamic Congruence Class Model (DCCM). The CCM as well as the DCCM, allow investigators to generate networks consistent with a broad class of probability distributions on essential network properties.  Below we review the key concepts of the CCM.

The CCM partitions the space of graphs with $N$ nodes, $\mathcal{G_N}$, into congruence classes, i.e., all graphs in a partition have the same values of essential network properties. The probability distribution on $\mathcal{G_N}$ for the CCM requires specification of $P_{\mathcal{C}}$, the probability mass function for the congruence classes defined by the essential network properties; as mentioned above,  $P_{\mathcal{C}}(\eta(g))$ is the total probability of all networks that are elements in $c_{\eta(g)}$, i.e., 

\begin{equation} \label{eq:networkprob_static_pc_1}
P_{\mathcal{C}}(\eta(g)) = \sum_{g* \in c_{\eta(g)}} P_{\mathcal{G}}(g*).
\end{equation}

Since the congruence classes represent the partition of the space $\mathcal{G_N}$ based on essential network properties, two networks within a congruence class must have the same probabilities of being observed. Therefore, the probability distribution on $\mathcal{G_N}$ for the CCM is the following:

\begin{equation} \label{eq:networkprob_static}
P_{\mathcal{G}}(g) \: = \: \left(\frac{1}{\vert c_{\eta(g)} \vert} \right) P_{\mathcal{C}}(\eta(g)),
\end{equation}

\noindent where $\vert c_{\eta(g)} \vert$ denotes the number of networks with essential property values equal to $\eta(g)$. 

The flexibility of the CCM results from the fact that the investigator can choose the  probability mass function on congruence classes, $P_{\mathcal{C}}$. The CCM allows  a broad range of models, including both parametric and nonparametric, in the assignment of  these probability mass functions. 

\subsection{Dynamic Congruent Class Model}\label{DCCM}

In the DCCM, the congruence classes are defined by both the static and dynamic essential network properties; by contrast, the CCM is based only on the former.  We denote $c_x$ as $c_{y,z}$ when it is necessary to separate the vector of values for the static and dynamic essential network properties as shown in Equation~\ref{eq:congruenc_class_def}. 

\begin{align} \label{eq:congruenc_class_def}
c_{y,z \vert g_{t-1}, \cdots, g_{t-k} } = \{ g_t : & \eta_s(g_t) = y, \nonumber \\
& \eta_d(g_t \vert g_{t-1} \cdots, g_{t-k}) = z \mbox{, and } g_t \in \mathcal{G}_t\}.
\end{align}

\noindent Adapting equation (\ref{eq:networkprob_static}) for the DCCM, the probability mass function on the space $\mathcal{G}_t$, is the following: 

\begin{equation} \label{eq:networkprob_dynamic}
P_{\mathcal{G}_t}(g_{t} \vert g_{t-k}, \cdots g_{t-n}) \: \propto \: \left(\frac{1}{\vert c_{\eta(g_{t} \vert g_{t-1}, \cdots g_{t-k})} \vert} \right) \times P_{\mathcal{C}_t}(\eta(g_t \vert g_{t-1}, \cdots, g_{t-k})).
\end{equation}

In the following sections, the congruence class of a network $g_t$ is restricted so that it only depends on the previous network, $g_{t-1}$. Therefore, the probability mass function in equation (\ref{eq:networkprob_dynamic}) simplifies to the following:

\begin{equation} \label{eq:networkprob_dynamic_Markov}
P_{\mathcal{G}_t}(g_{t} \vert g_{t-1}) \: = \: \left(\frac{1}{\vert c_{\eta(g_{t} \vert g_{t-1})} \vert} \right) \times P_{\mathcal{C}_t}(\eta(g_{t} \vert g_{t-1})).
\end{equation}
\noindent The decision to assume a Markov (or some other) process in defining the dynamic essential network properties does not restrict the collection of networks used to estimate the probability mass function on congruence classes, $P_{\mathcal{C}_t}$.  This flexibility allows the model to incorporate long-term and seasonal trends as well as degrees of uncertainty that vary over time based on the historical data.

As closed form expressions for $\vert c_{\eta(g_{t} \vert g_{t-1})} \vert$ are not available, sampling from $\mathcal{G}_t$ according to the probability mass function in equation (\ref{eq:networkprob_dynamic_Markov}) is performed by using a Metropolis-Hastings algorithm (MH)--a type of Markov Chain Monte Carlo (MCMC) procedure. To generate the  network at the $t^{th}$ step, $g_t$,  the algorithm starts by proposing a network, $gp_{t}$, based on the current state of the MCMC algorithm, denoted as $g_{t}'$, by toggling the existence of an edge (on or off) in $g_{t}'$. If the proposed network is accepted, based on equation (\ref{eq:prob_accept_DCCM}) below, the algorithm sets the current state $g_{t}' = gp_t$ and uses $gp_t$ as the basis of the next proposal; otherwise it remains on the current state and uses $g_{t}'$ as the basis of the next proposal. The algorithm continues for a set number of proposals and the final element of the chain is assigned to $g_t$. The algorithm produces an irreducible Markov chain among all graphs in $\mathcal{G}_t$. The equation for the acceptance probability for the MH algorithm is the following:

\begin{equation} \label{eq:prob_accept_DCCM}
P(\mbox{Accept } gp_{t} \vert g_t', g_{t-1}) = min \left(1, \frac{f(c_{\eta(gp_{t} \vert g_{t-1})}, c_{\eta(g_{t}' \vert g_{t-1})})}{f(c_{\eta(g_{t}' \vert g_{t-1})}, c_{\eta(gp_{t} \vert g_{t-1})})} \times \frac{P_{\mathcal{C}_t}(c_{\eta(gp_{t} \vert g_{t-1})})} {P_{\mathcal{C}_t}(c_{\eta(g_{t}' \vert g_{t-1})})} \right),
\end{equation} 

\noindent where $f(c_x, c_{x'})$ is the average number of elements in $c_{x'}$ that are valid proposals from an element $g \in c_x$. The non-standard acceptance probability formula in the MH algorithm arises because the DCCM's focus on congruence classes. Equation (\ref{eq:prob_accept_DCCM}) is identical to the acceptance probability derived in \cite{GBD13} except for the modification to the definition of the congruence classes that permits inclusion of dynamic essential network properties. 

\section{Comparison with (S)TERGMs}

\subsection{Theory}

Although TERGMs share certain features with the proposed DCCM, there is an important difference that makes DCCM particularly useful for network prediction--the ability to specify the functional form of network properties. In order to illustrate the distinction between TERGMs and DCCMs, it is helpful to review the formulation of TERGMs. Equation ~\ref{eq:TERGM} shows the PMF for TERGMs.

\begin{equation} \label{eq:TERGM}
P_{\mathcal{G}}(G = g_{t} | \omega) = \frac{1}{Z(\omega, g_{t-1},\cdots,g_{t-k})} \exp(\omega^T \eta(g_t, g_{t-1},\cdots,g_{t-k})),
\end{equation}

\noindent where $\omega$ is a vector of model parameters and $Z(\omega, g_{t-1},\cdots,g_{t-k})$ is a normalizing constant given the model parameters and previous networks; similar to the DCCM, $\eta(g_t, \cdots, g_1)$ is a function mapping networks $\{g_t,\cdots,g_1\}$ to real values of network properties. A TERGM (when nodal covariates are nominal) can be formulated as a DCCM by specifying the following for $P_{\mathcal{C}_t}(\eta(g_t \vert g_{t-1}, \cdots, g_{t-k}))$: 

\begin{equation} \label{eq:TERGMasDCCM}
P_{\mathcal{C}_t}(\eta(g_t \vert g_{t-1}, \cdots, g_{t-k})) = \frac{\vert c_{\eta(g_{t} \vert g_{t-1}, \cdots g_{t-k})} \vert}{Z(\omega, g_{t-1},\cdots,g_{t-k})} \exp(\omega^T \eta(g_t, g_{t-1},\cdots,g_{t-k})).
\end{equation}

\noindent Therefore, TERGMs restrict $P_{\mathcal{C}_t}$ to a specific functional form; by contrast, the functional form for DCCMs is unrestricted. The form associated with TERGMs arises by maximizing the entropy subject to constraints that specify only the mean of each network property \citep{MN10}. Therefore, when other constraints (such as limits on the variance of network properties) are present in the system being modelled by the dynamic network, TERGMs can be a poor fit for the data. In addition, the formulation of TERGMs results in a one-to-one correspondence between network properties and sufficient statistics, i.e., there is one parameter for each network property; this correspondence is not necessary for DCCMs.

STERGMs are a subclass of TERGMs that separates the formation and dissolution of edges in the dynamic network process. The primary motivation for developing STERGMs, as described by \citet{krivitsky2013separable}, is that the process for edge formation may differ from that for edge dissolution; this approach is advantageous in that one can postulate explicit models for both processes.

\subsection{Simulation}

We illustrate the difference between STERGMs and DCCMs through a simple example. Initialize $g_1$ as an Erd\H{o}s-R\'{e}nyi random graph with a population of 500 and a density of $1/500$. Generate $g_t$ by deleting and forming $nd_t$ and $nf_t$ edges, respectively, at random from $g_{t-1}$, where $nd_t, nf_t \sim Poisson(\lambda)$. Based on this process, we simulate a dynamic network of length 20, denoted as $g_{obs}^{\lambda} = \{g^{\lambda}_1,\cdots,g^{\lambda}_{20}\}$, for a specific value of $\lambda$. We assume that this dynamic network is observed by investigators and that the goal is to predict the next 100 future graphs based on $g_{obs}^{\lambda}$; we evaluate the usefulness of the STERGM and DCCM frameworks for making this prediction. 

Based on this network generation process the essential network properties are the number of edges that form and dissolve at each time step. Equivalently, we can specify the essential network properties as network density of graph $g_t$ and the number of edges that did not change between time steps $t-1$ and $t$, referred to as stability in \citet{hanneke2010discrete}. These two sets of essential network properties are equivalent because if one knows the number of number of edges that formed and dissolved between time steps $t-1$ and $t$, one can calculate the network density of graph $g_t$ and stability given $g_{t-1}$; the converse is also true.  

To specify a STERGM, we use the former specification for the essential network properties. As STERGMs impose a one-to-one correspondence between network properties and sufficient statistics, the only required sufficient statistic for the formation process is the number of edges that form between times $t-1$ and $t$. Similarly, for the dissolution process, the only require sufficient statistic is the number of edges that dissolved between times $t-1$ and $t$. No other modeling choice is necessary to specify a STERGM. To fit a STERGM, we use the `tergm' package in R \citep{krivitsky2019_tergm_R}. 

To specify a DCCM, we use the later specification described above for the essential network properties. DCCMs also require a functional form of the essential network properties, $P_{\mathcal{C}_t}(\eta(g_t \vert g_{t-1}, \cdots, g_{t-k}))$, which is chosen by the investigator; this step is the critical distinction between DCCMs and STERGMs. Any form for $P_{\mathcal{C}_t}$ can be specified (such as one based on the Poisson distribution or STERGMs), but for this illustration, we assume $P_{\mathcal{C}_t}$ follows a multivariate normal distribution. To fit a DCCM, we need to estimate the parameters of $P_{\mathcal{C}_t}(\eta(g_t \vert g_{t-1}, \cdots, g_{t-k}))$; we do so in this illustration by having the the mean and variance of the multivariate normal equal to the observed mean and variance across the graphs in $g_{obs}^{\lambda}$. We set the covariance to $0$, which is consistent with the generative process. 

For the fitted STERGM and DCCM, we generated a dynamic network of length 200 that represents times $\{t=21,\cdots,t=220\}$. Over these 200 predicted graphs, we compute the mean and variance for network density and stability. We repeated the process of generating a dynamic network 20 times, yielding have 20 dynamic networks of length 200 for both the fitted STERGM and DCCM for a given value $\lambda$.  Finally, we set  $\lambda$ to the following values: $1,5,10,20,30,40,50,60,70,80,90,$ and $100$. 

\begin{figure}
\centerline{\includegraphics[scale=0.75]{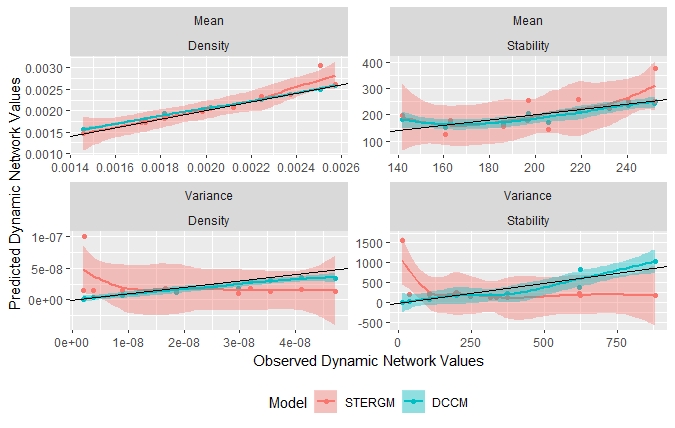}}
\caption{Simulation results for the observed graphs as well as those predicted by STERMs and DCCMs. The top left panel shows scatterplots for the mean network density averaged over all the 20 replications for a given value of $\lambda$;  that is, each point represents a distinct value of $\lambda$. The x-axis is the mean value for the observed graphs and the y-axis is the average mean value for the predicted graphs generated by the DCCM (blue points) and STERGM (red points). The bottom left panel shows scatterplots for the average variance associated with the observed (x-axis) and predicted (y-axis) dynamic networks for the network density statistic stratified by network generation method and $\lambda$. The right side shows similar information as the left side for the stability statistic.}
\label{fig:stergm}
\end{figure}

Figure~\ref{fig:stergm} shows the simulation results for the observed graphs as well as those predicted by STERMs and DCCMs. The top left panel shows scatterplots for the mean network density averaged over all the 20 replications for a given value of $\lambda$;  that is, each point represents a distinct value of $\lambda$. The x-axis is the mean value for the observed graphs and the y-axis is the average mean value for the predicted graphs generated by the DCCM (blue points) and STERGM (red points). The bottom left panel shows scatterplots for the average variance associated with the observed (x-axis) and predicted (y-axis) dynamic networks for the network density statistic stratified by network generation method and $\lambda$. The right side shows similar information as the left side for the stability statistic. 

Both DCCM and STERGM produced dynamic networks for which predicted mean network density (top left) and stability (top right) matched the observed values; this is illustrated by the fact that scatterplots followed closely the $x=y$ line. Nonetheless, the mean stability value for dynamic networks generated by STERGMs are further from the $x=y$ line compared to those generated by DCCMs. For the variance of network density and stability (bottom panels), DCCMs produced dynamic networks with values that are closely correlated with those observed. STERGMs, however, do not appear to capture the observed variances associated with $g_{obs}^{\lambda}$ for the network properites of network density and stability; this is due to the fact that the functional form of STERGMs does not constrain the simulated networks to have a variance that matches the observed variance. 

Next we compare the runtime performance of the DCCM and STERGM. In the illustration, the DCCM has to estimate the mean and variance of a multivariate normal distribution; however, more complex distributions could have been selected with longer estimation times. In addition, the estimation time for STERGMs depend heavily on the length of the observed dynamic network; we limited this length to 20 due to long estimation runtime of STERGMs for greater lengths. For these reasons, we focus on the time to generate a dynamic network of length 100 after the model has been fit. At each time step, DCCMs sample a single network, however STERGMs sample two networks (a formation and a dissolution network) and then combine them. Both the DCCM and STERGMs draw these networks using a MCMC algorithm. If we set the burnin as 5,000,000 steps for each of these networks, then DCCMs and STERGMs generate a single dynamic network of length 100 in approximately 318 and 285 seconds, respectively, based on a Intel Core i7-4600U CPU @ 2.10GHz. 

This simple example illustrates the advantage of the DCCM in allowing a flexible approach (parametric and non-parametric) for specifying the functional form of network properties,   $P_{\mathcal{C}_t}(\eta(g_t \vert g_{t-1}, \cdots, g_{t-k}))$--shown in Equation~\ref{eq:networkprob_dynamic}. Therefore, the DCCM models the distribution--and therefore the variance--associated with network properties in the observed graphs in models of the predicted graphs; this capability is further demonstrated in Section 7 for a more complex dynamic network. 

\section{Value of Predicting Networks: Simulation Studies}\label{Simulation}

The usefulness of the proposed approach lies in  its ability to predict  networks--not simply  collections of network properties.  In this section, we demonstrate the value of generating collections of predicted dynamic networks for the evaluation of interventions intended to modify network topology. The two simulation studies we discuss show that the association between dynamic network properties and outcomes can be complex even in  simple settings.  Both studies present interventions that are focused on reducing the spread of an infectious disease. However, the examples are general enough to represent complex systems across many settings.  

\subsection{Simulation Study 1}

The simulation study in this section mimics interventions intended to decrease the number of contacts during an epidemic of a communicable disease; the simulations use a simple susceptible-infected (SI) epidemic model for disease spread.  Each simulation models a population of 1000, where initially five individuals are infected with the disease.  At each time step individuals form new contacts and dissolve existing contacts; those infected may spread the disease to uninfected contacts.  The DCCM is used to control the formation and dissolution of contacts.  The model used a single static essential network property, number of edges, and a single dynamic essential network property, stability. We modeled $P_{\mathcal{C}_t}$ using a multivariate normal.

Six interventions are investigated in 6 simulations; the only difference among them is the rate at which the number of contacts in the population decreases. A seventh simulation, in which  the mean number of contacts does not decrease, represents  the absence of an intervention. At the start of each simulation, the contact networks have a mean of 1500 edges.  At the end, the mean number of edges for each of the six intervention simulations are 0, 250, 500, 750, 1000, and 1250; for the seventh, it remains at 1500 edges.  For all simulations, an average of 90\% of the edges persisted between consecutive networks.  The variance for the number of edges was based on the assumption that each had an equal probability of forming; for the  number of edges persisting variance was based on assuming  all edges had an equal probability of dissolving.   

\begin{figure}
\centerline{\includegraphics[scale=0.5]{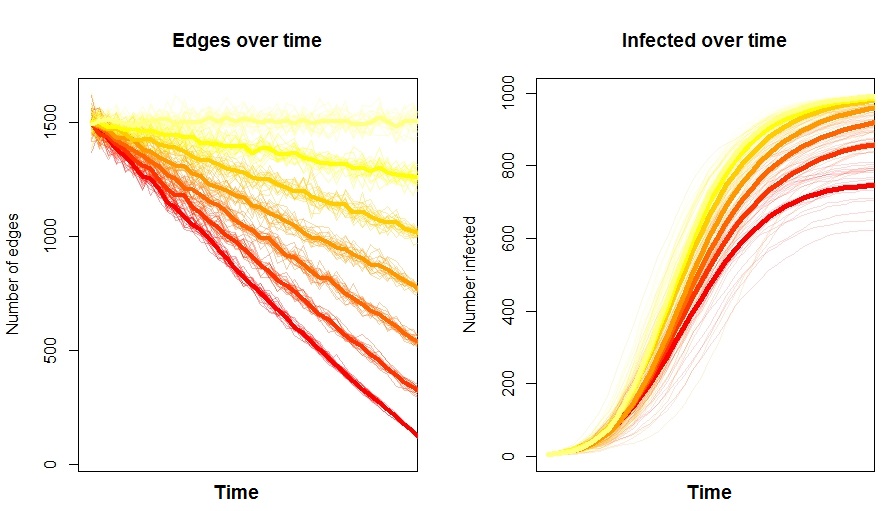}}
\caption{\textbf{Example 1.} The thin lines in the left panel of figure \ref{Simulation_1} show the number of edges over time for all of the simulations; each of the seven interventions was simulated 20 times; the seven thick lines show the average number of edges over time for each of the interventions. The thin lines in the right panel of figure \ref{Simulation_1} show the number infected over time for all of the simulations; similarly, the seven thick lines show the average number infected over time for each of the interventions.}
\label{Simulation_1}
\end{figure}

The thin lines in the left panel of figure \ref{Simulation_1} show the number of edges over time for all of the simulations; each of the seven settings was simulated 20 times. The seven thick lines show the average number of edges over time for each of the settings. For each setting, the average line and the variability around it indicate that the proposed method is performing as expected in  modeling the static essential network property; diagnostic plots (not shown) demonstrate that the proposed method also is correctly modeling the dynamic essential network property. The thin lines in the right panel of figure \ref{Simulation_1} show the number infected over time for all of the simulations; similarly, the seven thick lines show the average number infected over time for each of the settings. 

The association between the essential network properties and the number infected does not lend itself to characterization by a precise mathematical relationship. The curves shown in the right panel all have a slightly different shape, which makes it difficult to specify a precise mathematical relationship between the network topology and the cumulative infected over time. Simulations of this type, however, would permit an investigator to assess the potential impact of the six interventions by comparing the results of each of them (settings 1-6) to the results for the absence of the intervention (setting 7). This comparison is possible because of our ability to generate entire networks.

\subsection{Simulation Study 2}

This section describes simulation of interventions intended to control a communicable disease epidemic by decreasing the cumulative number of contacts while keeping constant the total length of time in relationships.  This study  setup is similar to that above: it models a population of 1000, of whom  five individuals are initially infected, assumes the same essential network properties, and uses an SI epidemic model to simulate the disease spread.  

\begin{figure}
\centerline{\includegraphics[scale=0.5]{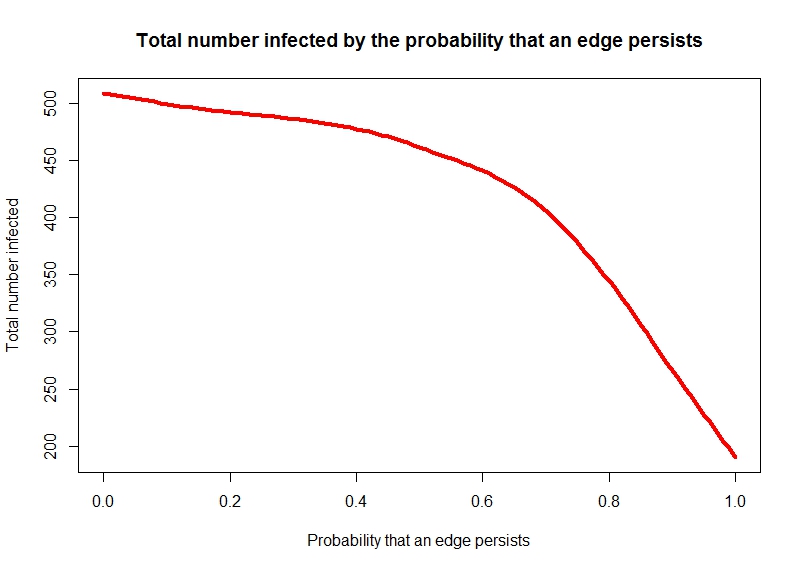}}
\caption{\textbf{Example 2.} The total number infected for each simulation for varying values of the probability that an edge persists between two time points.}
\label{Simulation_2}
\end{figure}

This study interventions impacts  the probability that an edge persists between two time points; this probability ranges from 0 to 1. We assume that in the absence of  intervention, the probability is zero. Throughout each simulation, the dynamic network has a mean of 800 edges; the variance for the essential network properties was based on the same assumptions as in the previous simulation study.  Figure \ref{Simulation_2} depicts the total number infected for each simulation for varying values of the probability that an edge persists between two time points.  As above,  it would be difficult to derive a precise mathematical description of the relationship between these two quantities. However, the simulation study permits an investigator to  assess the potential impact of an intervention designed to  decrease the cumulative number of contacts by comparing the results for each of the simulated interventions to the that representing the absence of the intervention. Once again, this comparison is made possible by the ability to generate entire networks.       

\section{Senate Bills 2003-2016}\label{SenateBills}

The longitudinal  network data represent relationships between US Senators as derived from bills introduced during the Unites States $108^{th}-114^{th}$ Senate. Each bill introduced in the US Senate has a single senator who serves as the sponsor of the bill; other senators may be associated with the bill as co-sponsors. A co-sponsorship network for month $t$ is generated by forming an undirected edge between the sponsoring senator and each of the co-sponsoring senators for bills introduced in month $t$. Prior research on the association between the topology of the co-sponsorship network and key legislation outcomes \citep{kirkland2014measurement}. \citet{fowler2006connecting} and \citet{kirkland2011relational} showed that network measures can predict a legislator's success. In addition, the overall topology of the network is associated with the amount of legislation that Congress passes  \citep{tam2010legislative}. 

We use the bills introduced during the $108^{th}-113^{th}$ Senate to predict the networks during the $114^{th}$ Senate. The next three subsections follow the conceptual framework outlined in section \ref{ConceptualFramework} and figure~\ref{figCF}. Sections 7.1 and 7.2 identify and predict the essential network properties; section 7.3 generates networks based on predicted estimates from the model developed in section 7.2. Section 7.4 investigates the GOF of the model to assess whether the modeled properties are sufficient to characterize the data. Section 7.5 demonstrates the usefulness of the method in predicting hypothetical effects of an intervention.

\subsection{Identifying essential network properties}

A salient feature in formation of collaborations among US Senators is party affiliation \citep{hanneke2010discrete}. We model three static essential network properties that capture mixing patterns between the two major political parties Democratic and Republican; senators designated as independent or socialist were assigned as Democrats. As the total number of senators fluctuated over the time intervals (e.g., Illinois only had one senator during December 2008) as did  the number affiliated with each political party,  we model the properties in a way that is  compatible with these data. The static essential network properties we model are the average number of edges that link senators according to party affiliation as defined below: 

\begin{align} \label{eq:networkmixing_sufficient}
\eta_s^{DD}(g_t) &= \vert \{E_{ij} : E_{ij} \in g_t, \DDg{m}{i}{}{g_t} = \mbox{D}, \mbox{ and } \DDg{m}{j}{}{g_t} = \mbox{D}\} \vert / \DDg{M}{D}{}{g_t} \\
\eta_s^{DR}(g_t) &= \vert \{E_{ij} : E_{ij} \in g_t, \DDg{m}{i}{}{g_t} = \mbox{D}, \mbox{ and } \DDg{m}{j}{}{g_t} = \mbox{R}\} \vert / \DDg{M}{D}{}{g_t} \\
\eta_s^{RR}(g_t) &= \vert \{E_{ij} : E_{ij} \in g_t, \DDg{m}{i}{}{g_t} = \mbox{R}, \mbox{ and } \DDg{m}{j}{}{g_t} = \mbox{R}\} \vert / \DDg{M}{R}{}{g_t},
\end{align}

\noindent where vector $\DDg{m}{}{}{g_t}$ represents the political affiliation for each node in $g_t$, i.e., $\DDg{m}{i}{}{g_t} \in \{D,R\}$ is the political affiliation for node $i$ (D for Democrat and R for Republican), and $\DDg{M}{D}{}{g_t}$ and $\DDg{M}{R}{}{g_t}$ are the number of Democrat and Republican senators, respectively, in network $g_t$. Let $\eta_s(g_t) = (\eta_s^{DD}(g_t), \eta_s^{DR}(g_t), \eta_s^{RR}(g_t))$. 

The vector for dynamic essential network properties, $\eta_d(g_t \vert g_{t-1})$, describes the stability between $g_{t}$ and $g_{t-1}$ for each pair of political affiliations, (D,D), (D,R), and (R,R); the reason for specifying three dynamic essential network properties as opposed to one is to avoid ``churn'', as described in \citet{krivitsky2013separable}. The term $\eta_d(g_t \vert g_{t-1})$ consists of the average number of common edges in $g_t$ and $g_{t-1}$ that link: 1) a Democrat to another Democrat, 2) Democrat to a Republican, and 3) Republican to another Republican and denoted as $\eta^{DD}_d(g_t \vert g_{t-1})$, $\eta^{DR}_d(g_t \vert g_{t-1})$, and $\eta^{RR}_d(g_t \vert g_{t-1})$, respectively. The formulas for dynamic essential network properties are presented below:

\begin{align} \label{eq:networkmixing_stab_2}
\eta^{DD}_d(g_t \vert g_{t-1}) = &\vert \{E_{ij} : E_{ij} \in g_t, E_{ij} \in g_{t-1}, \DD{m}{i}{} = \mbox{D}, \mbox{ and } \DD{m}{j}{} = \mbox{D}\} \vert / \DDg{M}{D}{}{g_t} \\
\eta^{DR}_d(g_t \vert g_{t-1}) = &\vert \{E_{ij} : E_{ij} \in g_t, E_{ij} \in g_{t-1}, \DD{m}{i}{} = \mbox{D}, \mbox{ and } \DD{m}{j}{} = \mbox{R}\} \vert / \DDg{M}{D}{}{g_t} \\
\eta^{RR}_d(g_t \vert g_{t-1}) = &\vert \{E_{ij} : E_{ij} \in g_t, E_{ij} \in g_{t-1}, \DD{m}{i}{} = \mbox{R}, \mbox{ and } \DD{m}{j}{} = \mbox{R}\} \vert / \DDg{M}{R}{}{g_t}.
\end{align}

\begin{figure}
\centerline{\includegraphics[scale=0.35]{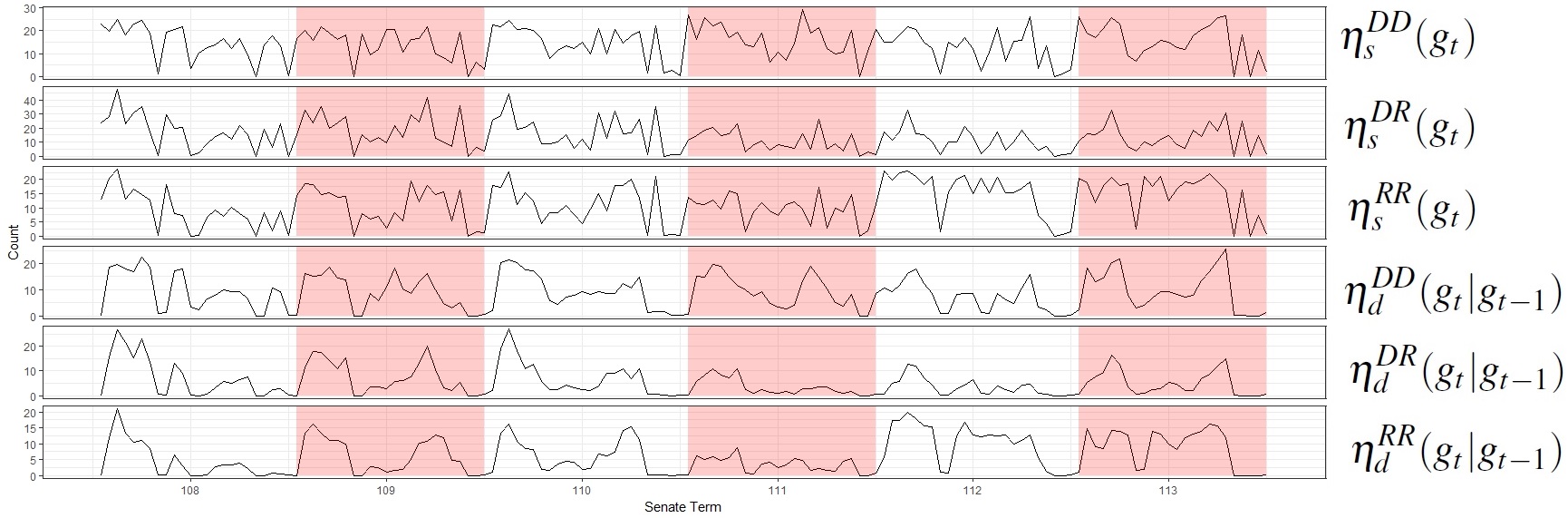}}
\caption{\textbf{US Senate Network Statistics.} The black lines depict the values of $\eta(g_t \vert g_{t-1}) = \{\eta_s^{DD}(g_t), \eta_s^{DR}(g_t),\eta_s^{RR}(g_t), \eta^{DD}_d(g_t \vert g_{t-1}) , \eta^{DR}_d(g_t \vert g_{t-1}) , \eta^{RR}_d(g_t \vert g_{t-1}) \}$ for the $108-113^{th}$ Senate. The shaded sections represent the $109^{th}$,  $111^{th}$, and $113^{th}$ Senates, while the non-shaded sections represent the $108^{th}$, $110^{th}$, and $112^{th}$.}
\label{Senate_Bill_Mixing}
\end{figure}

\noindent Let $\eta(g_t \vert g_{t-1}) = \{\eta_s^{DD}(g_t), \eta_s^{DR}(g_t),\eta_s^{RR}(g_t), \eta^{DD}_d(g_t \vert g_{t-1}) , \eta^{DR}_d(g_t \vert g_{t-1}), \eta^{RR}_d(g_t \vert g_{t-1}) \}$. The black lines in the top three plots of figure \ref{Senate_Bill_Mixing} depict the values of $\eta_s$, while the bottom three plots depict $\eta_d$ for the $108^{th}-113^{th}$ Senates. We excluded the dynamic essential network property values where the months $t-1$ and $t$ are associated with different senate terms, and set the value to zero in figure \ref{Senate_Bill_Mixing} to retain the same time scale as the static essential network properties. 

\subsection{Predicting Network Statistics}

We  develop a model to predict $\eta(g_t)$ for the $114^{th}$ Senate using  data from the $108^{th}-113^{th}$ Senates. The prediction model is used to specify $P_{\mathcal{C}_t}$ for $t \in \{\mbox{January} \: 2015,\cdots, \mbox{December} \: 2016\}$. Let $X(t)$ be the vector of random variables associated with the static and dynamic essential network properties; $X(t)$ is comprised of the three random variables for the static properties, denoted as $Y_s^{DD}(t), Y_s^{DR}(t)$, and $Y_s^{RR}(t)$, and three random variables for the dynamic properties, denoted as $Z_d^{DD}(t), Z_d^{DR}(t)$, and $Z_d^{RR}(t)$. 

An advantage of the DCCM is that the development of the prediction model for $\eta(g_t)$ does not require the Markov assumption used in defining the dynamic essential network statistics; we  use all of the historical networks from the $108^{th}-113^{th}$ Senates, and  denote this collection as  $\vec{g_h}$. We base our predictions of each component of $X(t)$ on an autoregressive moving average (ARMA) model with a seasonal component in order to capture the periodic fluctuations of the network statistics associated with the congressional election cycle. The seasonal $ARMA(p,q)(P,D)_s$ model for $Y_s^{DD}(t)$ has the following form:

\begin{equation} \label{eq:ARMA}
\Phi(B^s)\phi(B)Y_s^{DD}(t) = \Theta(B^s)\theta(B)W_t,
\end{equation}

\noindent where 

\begin{align} \label{eq:ARMA_2}
\Phi(B) &= 1-\sum_{j=1}^P \Phi_j B^{js},\\
\phi(B) &= 1-\sum_{j=1}^p \phi_j B^{j},\\
\Theta(B^s) &= 1-\sum_{j=1}^Q \Theta_j B^{js},\\
\theta(B) &= 1-\sum_{j=1}^q \theta_j B^{j}, \\
BY_s^{DD}(t) &= Y_s^{DD}(t-1), \mbox{ and } \\
W_t &= Normal(0,\sigma^2).
\end{align}

A separate $ARMA(p=3,q=1)(P=2, Q=1)_{24}$ model--a model with p=3 autoregressive terms and q=1 moving-average term and a seasonal component with P=2, Q=1 and period of s=24 months--was used to model each of the static essential network properties, $Y_s^{DD}(t), Y_s^{DR}(t)$, and $Y_s^{RR}(t)$. For the dynamic essential network properties, $Z_d^{DD}(t), Z_d^{DR}(t)$, and $Z_d^{RR}(t)$, we fit the same model except the seasonal component had a period of s=23 months; we excluded the values where the months $t-1$ and $t$ are associated with different senate terms. In order to combine these separate models, we use the product distribution. Therefore, 

\begin{align} \label{eq:eta_predict_2}
P_{\mathcal{C}_t}(X(t) = (y_s^{DD}(t), y_s^{DR}(t), y_s^{RR}(t), z_d^{DD}(t), z_d^{DR}(t), z_d^{RR}(t))\vert \vec{g_h}) = \nonumber \\
\left\{ \begin{gathered}
\Pi_{k \in \{DD,DR,RR\}} P(Y_s^k(t) = y_s^{k}(t) \vert \vec{g_h})
\mbox{ if } t = \mbox{ January 2015} \hfill \\
\Pi_{k \in \{DD,DR,RR\}} P(Y_s^k(t) = y_s^{k}(t) \vert \vec{g_h}) * P(Z_d^k(t) = z_d^{k}(t) \vert \vec{g_h})
\mbox{ else,} \hfill
\end{gathered} \right.
\end{align}

\noindent where $P(Y_s^k(t) = y_s^{k}(t) \vert \vec{g_h})$ and $P(Z_d^k(t) = z_d^{k}(t) \vert \vec{g_h})$ are based on an $ARMA(p=3,q=1)(P=2, Q=1)_{24}$ and $ARMA(p=3,q=1)(P=2, Q=1)_{23}$ models, respectively, and $k \in \{DD, DR, RR\}$. The models were fit using the R package Forecast \citep{Rforecast}. Based on the ARMA model, the predicted distribution for each essential and stability network property follows a normal distribution. Therefore, $P_{\mathcal{C}_t}(X(t))$ can be represented as the following multivariate normal distribution: 

\begin{equation} \label{eq:eta_predict_3}
\begin{aligned}
P_{\mathcal{C}_t}(X(t)) = MVN(\mu_t, \Sigma_t). 
\end{aligned}
\end{equation}

\begin{figure}
\centerline{\includegraphics[scale=0.35]{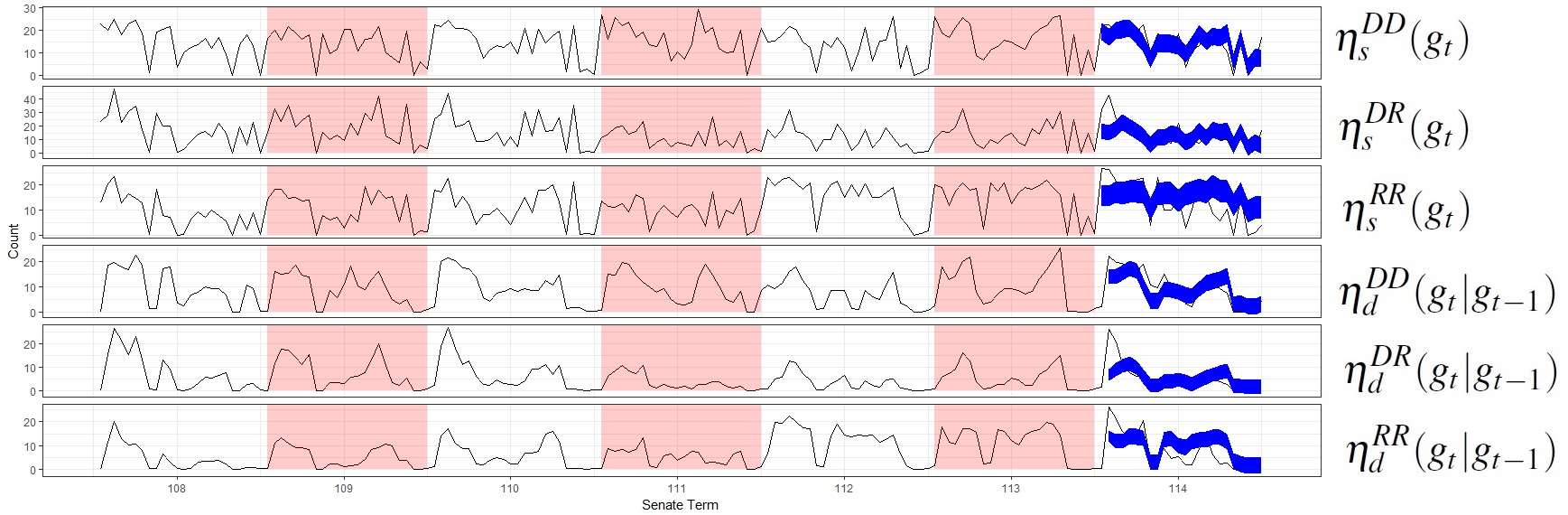}}
\caption{\textbf{US Senate Network Statistics.} The black lines depict the values of $\eta(g_t \vert g_{t-1}) = \{\eta_s^{DD}(g_t), \eta_s^{DR}(g_t),\eta_s^{RR}(g_t), \eta^{DD}_d(g_t \vert g_{t-1}) , \eta^{DR}_d(g_t \vert g_{t-1}) , \eta^{RR}_d(g_t \vert g_{t-1}) \}$ for the $108-114^{th}$ Senate. The shaded sections represent the $109^{th}$, $111^{th}$ and $113^{th}$ Senates, while the non-shaded sections represent the $108^{th}, 110^{th}$ and $112^{th}$ Senates. The areas defined by the blue regions represent the predicted intervals.}
\label{Senate_Bill_Mixing_Pred}
\end{figure}

For purposes of illustration, we set $\Sigma_t$ so that two standard deviations cover 50\% of the prediction interval; using a $\Sigma_t$ for which two standard deviations  cover 95\% of the prediction interval would have large uncertainty and therefore reduce the clarity of the figures (no modification of the method is required to use other standard deviations, such as ones which would cover 90\% or 95\% of the prediction interval).  In figure \ref{Senate_Bill_Mixing_Pred}, the areas defined by the blue regions represent the revised prediction intervals. 

As our framework places only minimal restrictions on the selection of model for $X(t)$, an investigator could select the most appropriate model, such as  a vector autoregressive (VAR)  or   non-Gaussian models without modification to the method.  We chose simple models for clarity of presentation. 

\subsection{Generation of Predicted Networks}

\subsubsection{Overview}

This section describes the generation of  networks that represent the predicted sponsor/co-sponsor relationships between senators for the $114^{th}$ Senate for the months January, 2015 to December, 2016. Using equation (\ref{eq:networkprob_dynamic_Markov}), the probability distribution for the predicted networks, $P_{\mathcal{G}_t}(g_{t} \vert g_{t-1})$, is the following: 

\begin{equation} \label{eq:prob_cc_networkmixing}
P_{\mathcal{G}_t}(G_t = g_{t} \vert g_{t-1}) \: = \: \left(\frac{1}{\vert c_{\eta(g_{t} \vert g_{t-1})} \vert} \right) \times 
P_{\mathcal{C}_t}(X(t) = \eta(g_{t} \vert g_{t-1})), 
\end{equation}

\noindent where $X(t) \sim MVN(\mu_t, \Sigma_t)$ is estimated in the previous section.

\subsubsection{Results}

The procedure described in the Appendix was used to generate dynamic networks predicting the evolution of the co-sponsor relationships for bills introduced from January 2015 to December 2016, i.e., the $114^{th}$ Senate. Each dynamic network is comprised of 24 static networks--one for each month. The procedure was repeated 500 times. Let $\vec{g^{pred}_i}$ denote the $i^{th}$ generated predicted dynamic network, where $\vec{g^{pred}_i}[t]$ represents the network at time $t \in \{$January 2015,$\cdots$, December 2016$\}$. To evaluate the procedure, the  predicted dynamics networks generated by our model are compared to the estimated probability distribution of essential network properties, i.e., $P_{\mathcal{C}_t}$, shown in equation (\ref{eq:eta_predict_3}). To conduct this evaluation, we calculate $\eta_s^{DD}(g_t), \eta_s^{DR}(g_t),\eta_s^{RR}(g_t), \eta^{DD}_d(g_t \vert g_{t-1}) , \eta^{DR}_d(g_t \vert g_{t-1}),$ and $\eta^{RR}_d(g_t \vert g_{t-1})$ for all of the generated predicted dynamic networks. Let $\vec{y_s^{DD}(t)}$ denote a vector for $\eta_s^{DD}$ at time $t$ for all predicted dynamic networks, $i \in \{1,\cdots,500\}$, i.e., 

\begin{equation} \label{eq:proj_prediction}
\vec{y_s^{DD}(t)} = \{\eta_s^{DD}(\vec{g^{pred}_1}[t]), \cdots, \eta_s^{DD}(\vec{g^{pred}_{500}}[t])\}.
\end{equation} 

\noindent Similarly, define $\vec{y_s^{DR}(t)}, \vec{y_s^{RR}(t)}, \vec{z_d^{DD}(t)}, \vec{z_d^{DR}(t)},$ and $\vec{z_d^{RR}(t)}$.

\begin{figure}
\centerline{\includegraphics[scale=0.35]{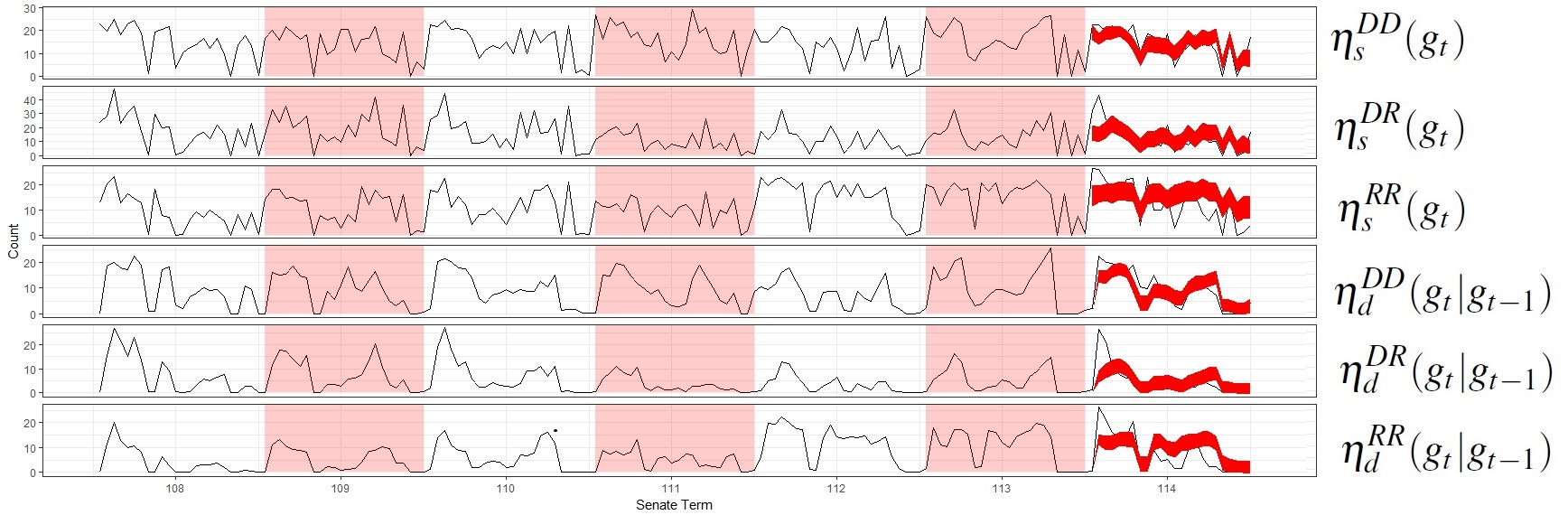}}
\caption{\textbf{US Senate Network Statistics.} The black lines depict the values of $\eta(g_t \vert g_{t-1}) = \{\eta_s^{DD}(g_t), \eta_s^{DR}(g_t),\eta_s^{RR}(g_t), \eta^{DD}_d(g_t \vert g_{t-1}) , \eta^{DR}_d(g_t \vert g_{t-1}) , \eta^{RR}_d(g_t \vert g_{t-1}) \}$ for the $108-114^{th}$ Senate. The shaded sections represent the $109^{th}$, $111^{th}$, and $113^{th}$ Senates, while the non-shaded sections represent the $108^{th}, 110^{th}$ and $112^{th}$ Senates. The areas defined by the red regions represent the 2.5\% and 97.5\% quantiles of $\eta$ applied to the predicted dynamic networks. }
\label{Senate_Bill_Mixing_Sim}
\end{figure}

The red region in the top plot of figure \ref{Senate_Bill_Mixing_Sim} represents 2.5\% and 97.5\% quantiles of $\vec{y_s^{DD}(t)}$ at each time point $t \in \{$January 2015,$\cdots$, December 2016$\}$. The five subsequent plots represent 2.5\% and 97.5\% quantiles of $\vec{y_s^{DR}(t)}, \vec{y_s^{RR}(t)}, \vec{z_d^{DD}(t)}, \vec{z_d^{DR}(t)},$ and $\vec{z_d^{RR}(t)}$, respectively. The blue regions in figure~\ref{Senate_Bill_Mixing_Pred}, display the 2.5\% and 97.5\% quantiles of the network property values based on the estimated probability distribution of essential network properties, i.e., $P_{\mathcal{C}_t}$. The red regions figure~\ref{Senate_Bill_Mixing_Sim} and blue regions in figure~\ref{Senate_Bill_Mixing_Pred} are nearly identical. Therefore, figure \ref{Senate_Bill_Mixing_Sim} provides evidence that networks generated by the proposed method  are appropriate; this result is expected as we were able to calculate $f(C_{g}, C_{h})$ exactly. See Appendix for details. 

\subsection{Goodness-of-Fit}

\citet{hanneke2010discrete} proposed an extension of the approach by \citet{hunter2008goodness} to evaluate goodness-of-fit heuristically; we use their method to assess the fit of the predicted networks. \citet{hunter2008goodness} and \citet{hanneke2010discrete} used the same networks to build a model and assess its fit. The general approach is to compare values of network properties that were not included in the model between the true and model-generated networks; this approach can be sensitive to the network properties chosen as there is no set of ``basis" network properties that can guarantee that one has identified the complete set of essential properties. This approach shares similarities with the network classification method by \citet{airoldi2011network}. We visually inspect the differences, but statistical tests can be applied, such as a Chi-squared test if the values can be binned. As our focus is on forecasting networks, there are challenges in assessing goodness-of-fit. In our analysis, however, the predicted $114^{th}$ Senate sponsorship networks are actually fully observed, but were excluded from our modeling; therefore, we are able to base our GOF on a comparison of the true networks to those we predicted. We consider the approach particularly useful since poor fit can arise from either 1) important essential network properties that are missing in the model or 2) a network structure of the $114^{th}$ Senate that is fundamentally different from the previous Senates.  

\begin{figure}
\centerline{\includegraphics[scale=0.35]{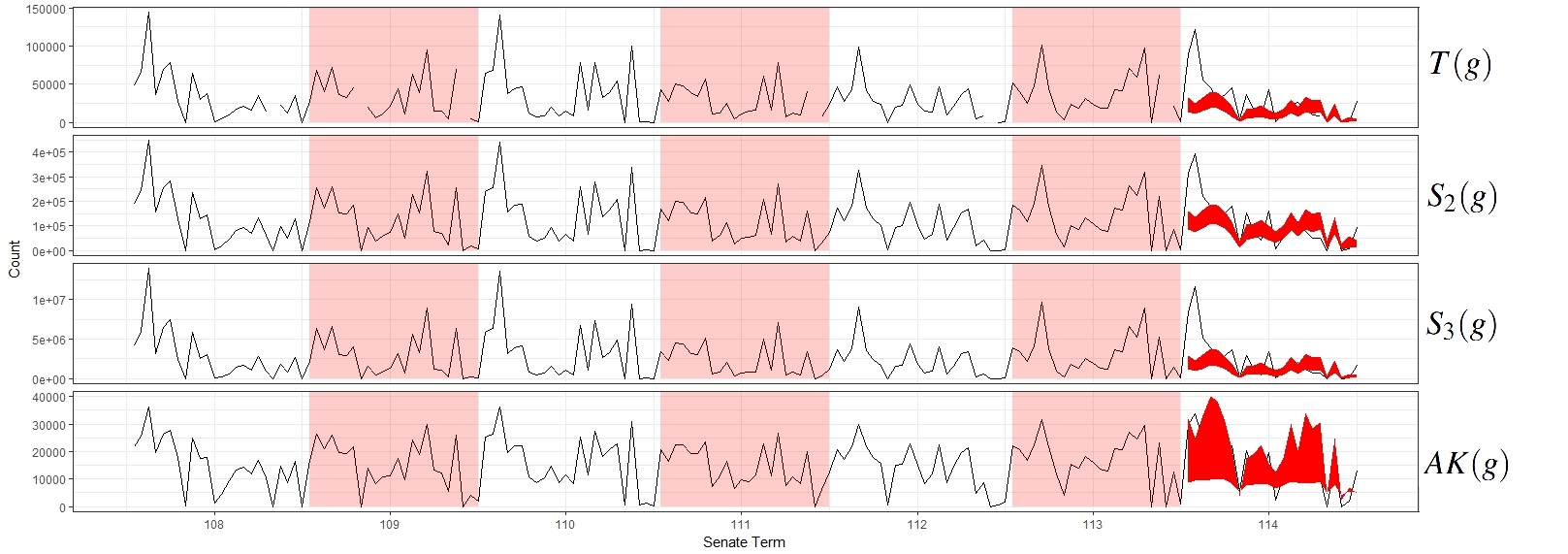}}
\caption{\textbf{Goodness-of-fit Plots.} The black lines depict the values for the $108-114^{th}$ Senate of the following network properties: number of triangles, number of 2-stars, number of 3-stars, and alternating k-stars. The blue regions represent the 2.5\%-97.5\% quantiles of network statistics calculated from the simulated predicted networks.}
\label{Senate_Bill_Mixing_GOF}
\end{figure}

Figure \ref{Senate_Bill_Mixing_GOF} shows the values of four additional network properties that were not explicitly modeled: number of triangles ($T(g)$), number of 2-stars ($S_2(g)$), number of 3-stars ($S_3(g)$), and alternating k-stars ($AK(g)$). The expressions for these four network properties are:

\begin{align}
T(g) &= \sum_{i,j,k \in g} I_{e_{ij} \in E(g)} I_{e_{ik} \in E(g)} I_{e_{jk} \in E(g)} \\
S_i(g) &= \sum_{i \in g} {\sum_{j \in g} I_{e_{ij} \in E(g)} \choose i} \\
AK(g) &= \sum_{i=2}^{i=n-1} (-1)^{i} S_i/\tau^{i-2}.
\end{align}

\noindent The black line in each plot of figure \ref{Senate_Bill_Mixing_GOF} depicts the observed values for a network property for the $108^{th}-114^{th}$ Senate. Each red region represents the 2.5\%-97.5\% quantiles of the network statistics calculated from the simulated predicted networks. The first plot provides the number of triangles;  the bottom three plots are related to degree distribution. The number of triangles, 2-stars and 3-star statistics from the simulated predicted networks appear to fit the observed network statistics closely, except for the early months of 2015. The number of relationships in the network during the early months of 2015 were higher than historical averages, which may indicate that the lack of fit was due to a change in the network structure for the $114^{th}$ Senate compared to prior terms.  We note that in 2015, control of the Senate shifted from Democratic to Republican for the first time since 2006.  There seems to be a good fit with the alternating k-star property.

\subsection{Simulation of proposed intervention}\label{Scenarios}

Several political analysts have proposed that the increasing use of gerrymandering at the state level to create congressional districts that favor the party in power have decreased bipartisanship \citep{enten538}.  \citet{enten538} states that ``Gerrymandering contributes to issues like the drop in competitive elections, extremism and gridlock, but it's far from their sole cause.'' He goes on to state: ``What's behind the disappearance of so many competitive districts? Gerrymandering is part of the story...It's clear that most redistricting schemes that ignore politics and race would yield more competitive U.S. House districts--i.e., those with a partisan lean of 10 percentage points or less--than we currently have.'' He also quotes John Kasich in his 2016 state address: ``Ideas and merits should be what wins elections, not gerrymandering. When pure politics is what drives these kinds of decisions, the result is polarization and division. I think we've had enough of that. Gerrymandering needs to be [in] the dust bin of history.'' While some states, such as California have anti-gerrymandering laws, there is no such federal law in the United States.

\begin{figure}
\centerline{\includegraphics[scale=0.5]{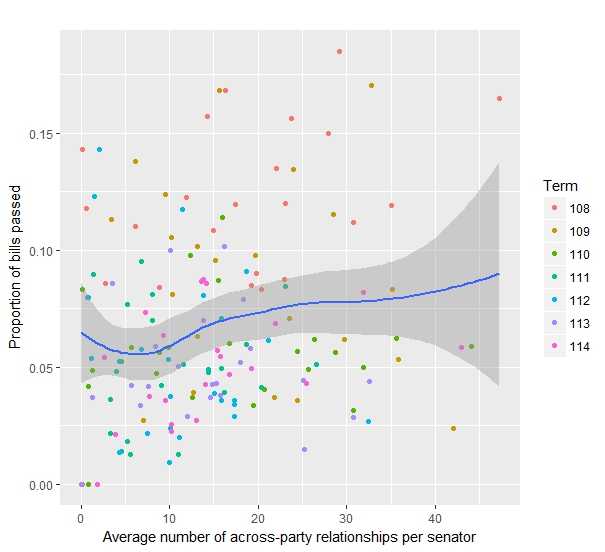}}
\caption{\textbf{Passage of Bills.} The $\eta_s^{DR}(g_t)$ values for the $108-114^{th}$ are shown on the x-axis; color coded by congressional term.  The y-axis shows the proportion of bills introduced in the Senate that passed the Senate by month. The blue line shows the loess curve.}
\label{Senate_Passing_DR}
\end{figure}

Political polarization, resulting in part from gerrymandering, has been proposed as a cause of congressional gridlock that has become the norm over the past several terms \citep{jacobson2016polarization, carson2007redistricting}. Degree of  bipartisanship can be measured as the level of bipartisan support of bills--specifically  the  number of  ties between senators of different parties in the sponsor/co-sponsor networks, i.e., values for $\eta_s^{DR}(g_t)$. Figure \ref{Senate_Passing_DR} shows a direct association between $\eta_s^{DR}(g_t)$ and the proportion of bills introduced in the Senate that passed the Senate (a univariable regression has a p-value of 0.0225).  

As passage of bills in the Senate is required for new laws and the operation of US government; it is of  interest to investigate consequences of decreased bipartisan co-sponsoring of bills on their probability of passage.  We consider three hypothetical scenarios starting at the beginning of the $115^{th}$ Senate.  The first assumes that the essential network properties for the $115^{th}$ Senate follow the prediction model shown in equation (\ref{eq:eta_predict_3}).  The second and third make the same assumption, except that the number of across-party relationships is reduced by 95\% (perhaps resulting from impact of factors like social media) and increased by 100\%. The comparison of these three scenarios provides an estimate of the impact on bill passing rates of increases or decreases in bipartisan support  compared to historically observed trends.  Alternative scenarios with varying parameter choices would be easy to investigate.  

Predicting  bill passage rates under these three scenarios requires two statistical models.  The first links covariates, including network properties, to the outcome of bill passage rates.  These rates may depend on lower level network properties, such as the amount of across-party relationships, as well as higher order properties, such as centrality measures.  In order to investigate this each month, we developed a basic random forest model using the network properties modeled in the previous section as well as variables for the number of components, size of the largest component, eigenvalue centrality, the maximum value for closeness and betweenness, and the number of individuals from each party. We use this model to illustrate the proposed framework and acknowledge that additional research and modeling is necessary to improve accuracy of  prediction. The node purity metric, which indicates the importance of a variable in a random forest model, shows that eigenvalue centrality has an importance similar to that of the amount of across-party relationships, which provides support for the notion that higher order network properties impact bill passage rates in the US Senate. This finding is in line with previous research that has demonstrated correlation of centrality measures of the sponsor/co-sponsor network  with the number of amendments and the associated bills that a senator will pass \citep{fowler2006connecting}.

\begin{figure}
\centerline{\includegraphics[scale=0.35]{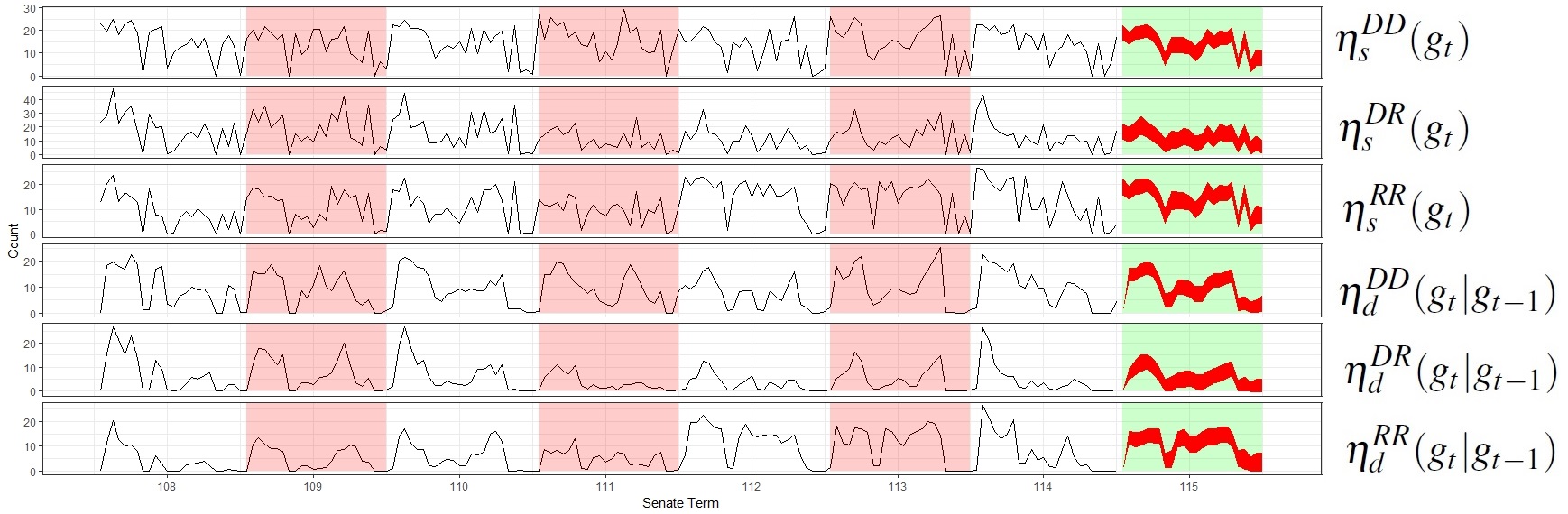}}
\caption{\textbf{Scenario 1.} The black lines depict the values of $\eta(g_t \vert g_{t-1}) = \{\eta_s^{DD}(g_t), \eta_s^{DR}(g_t),\eta_s^{RR}(g_t), \eta^{DD}_d(g_t \vert g_{t-1}) , \eta^{DR}_d(g_t \vert g_{t-1}) , \eta^{RR}_d(g_t \vert g_{t-1}) \}$ for the $108-114^{th}$ Senate. The red shaded sections represent the $109^{th}$, $111^{th}$ and $113^{th}$ Senates, while the non-shaded sections represent the $108^{th}, 110^{th}$ and $112^{th}$ Senates. The areas defined by the red regions on the green shaded section represent the predicted number of edges for the unobserved $115^{th}$ Senate under scenario 1.}
\label{Senate_Intervention_S1}
\end{figure}

\begin{figure}
\centerline{\includegraphics[scale=0.35]{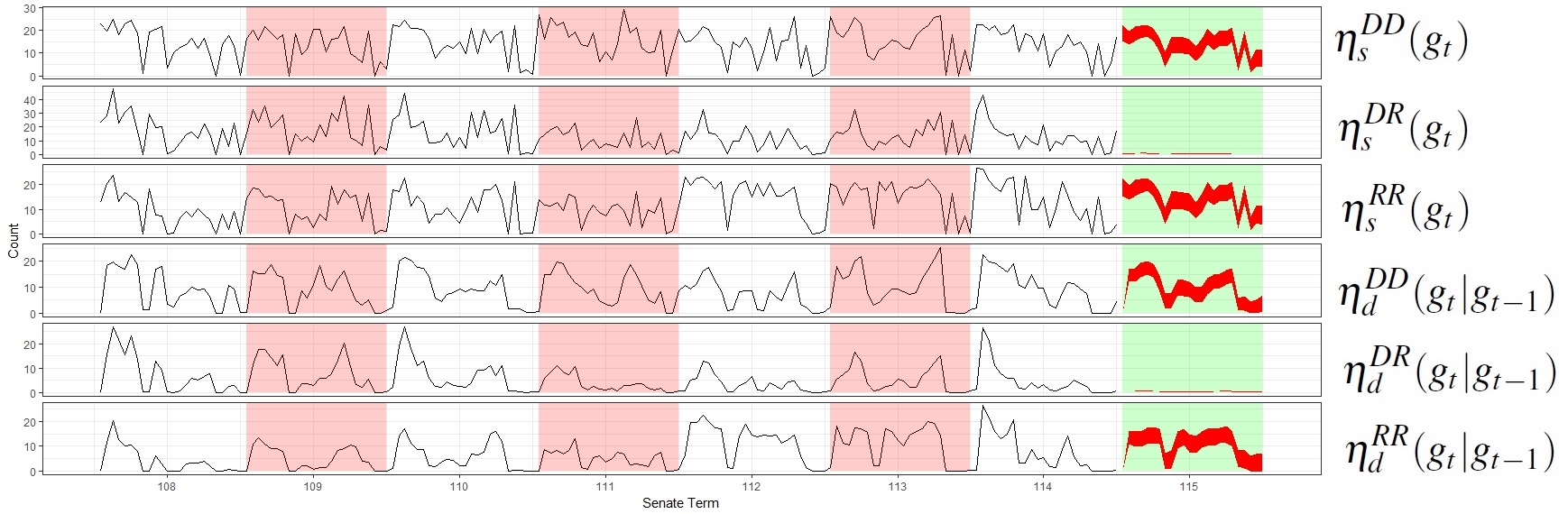}}
\caption{\textbf{Scenario 2.} The black lines depict the values of $\eta(g_t \vert g_{t-1}) = \{\eta_s^{DD}(g_t), \eta_s^{DR}(g_t),\eta_s^{RR}(g_t), \eta^{DD}_d(g_t \vert g_{t-1}) , \eta^{DR}_d(g_t \vert g_{t-1}) , \eta^{RR}_d(g_t \vert g_{t-1}) \}$ for the $108-114^{th}$ Senate. The red shaded sections represent those quantities for  $109^{th}$, $111^{th}$ and $113^{th}$ Senates, while the non-shaded sections represent them for $108^{th}, 110^{th}$ and $112^{th}$ Senates. The areas defined by the red regions on the green shaded section represent the predicted values for the unobserved $115^{th}$ Senate under scenario 2.}
\label{Senate_Intervention_S2}
\end{figure}

\begin{figure}
\centerline{\includegraphics[scale=0.35]{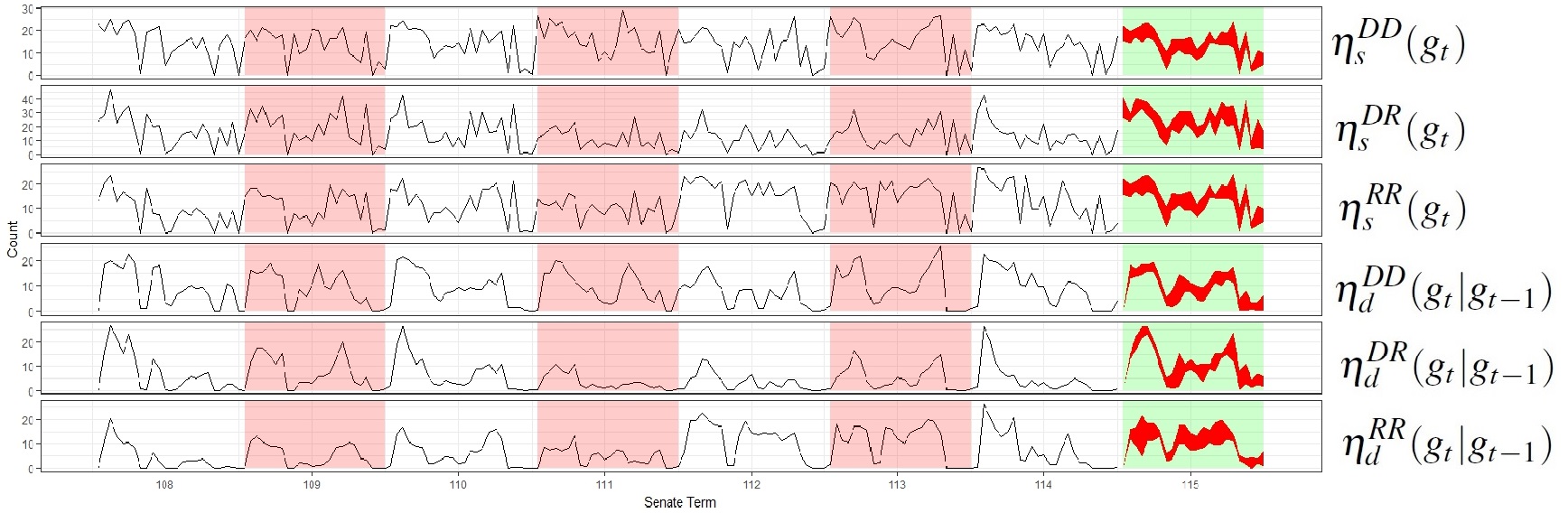}}
\caption{\textbf{Scenario 3.} The black lines depict the values of $\eta(g_t \vert g_{t-1}) = \{\eta_s^{DD}(g_t), \eta_s^{DR}(g_t),\eta_s^{RR}(g_t), \eta^{DD}_d(g_t \vert g_{t-1}) , \eta^{DR}_d(g_t \vert g_{t-1}) , \eta^{RR}_d(g_t \vert g_{t-1}) \}$ for the $108-114^{th}$ Senate. The red shaded sections represent those quantities for  $109^{th}$, $111^{th}$ and $113^{th}$ Senates, while the non-shaded sections represent them for $108^{th}, 110^{th}$ and $112^{th}$ Senates. The areas defined by the red regions on the green shaded section represent the predicted values for the unobserved $115^{th}$ Senate under scenario 3.}
\label{Senate_Intervention_S3}
\end{figure}

The second model predicts values of covariates that are included in the first model. Therefore, we need to predict values for the network properties described in equations (8) to (13) as well as the number of components, size of the largest component, eigenvalue centrality, and the maximum value for closeness and betweenness; the number of individuals from each party is based on counts at the start of the $115^{th}$ Senate. For all three scenarios, the predicted values for network properties described in equations (8) to (13) are either specified by the prediction model developed in section 6.2 or through the assumptions of the scenarios.  However, it is difficult to estimate the remaining properties. Therefore, we use our framework to generate networks and use them to estimate the remaining properties. Figures \ref{Senate_Intervention_S1}, \ref{Senate_Intervention_S2}, and \ref{Senate_Intervention_S3} show that the proposed method can generate networks under the three scenarios. Note, for the first scenario, is it possible to predict the remaining properties by developing a time series models for each property.  

Applying the first prediction model using the estimated covariates, we predict that  the average monthly pass rate would decline by 3.9\% in scenario 2 compared to scenario 1 and increase by 3.6\% for scenario 3 compared to scenario 1.  These results  imply a modest change in bills passing the Senate if bipartisan support erodes or increases faster than predicted according to historically observed trends.  

\section{Discussion}\label{Discussion}

The proposed framework for predicting dynamic network allows for flexible modeling of the joint distribution of essential network properties at time $t$ based on previously observed networks.  This flexibility permits the use of a broad class of approaches to model trends, seasonal variability, uncertainty, and changes in population composition. The flexibility makes the method particularly well suited to serve as a basis for designing potential interventions that modify network topology as investigators are able to model changes in network properties that result from interventions and compare these changes to those based on historical network trends--as in our illustration. 

In addition to the application we present, there are a range of research areas where the proposed method is applicable.  One is investigation of the impact of inventions, such as treatment and behavior changes, to mitigate the spread of diseases--for example in  investigating the  impact of reducing sexual partner concurrency to reduce the spread of HIV. Reducing concurrency is tantamount to reducing  the degree of individuals in a sexual contact network below 2 at a point in time.  Assessing the impact of concurrency is challenging because modifying one network property (degree in this example) will modify others as well, including higher order network properties \citep{goyal2015sampling}. The reason for this modification is that individuals who cease concurrent relationships may either form non-concurrent relationships or forgo relationships entirely for periods of time--either possibility will impact higher order network properties. The forming of other non-concurrent relationships may lead to changes in degree assortativity in the network (aggregated over time). High-order properties, in particular degree assortativity, have been shown to impact disease spread \citep{pastor2015epidemic}.  Therefore, as with the Senate example, it is necessary to generate the entire network and not just summary statistics; estimates of higher properties cannot be computed easily.  

The impact of uncertainty in network property estimates associated with dynamic networks has received little attention compared to other areas of network science.  However, the existence of sharp thresholds in relationships among properties for static networks has been well-documented \citep{ER60, WS98, MN10}.  Therefore is it possible for a small change in a given dynamic network property to have significant impact on processes operating on the network.  Further research is necessary to understand the impact of variability in network properties has for predicting intervention impacts on social systems. 

Additional methodological work is needed to evaluate the form of $f(C_{g}, C_{h})$ for additional network statistics. The CCM has been expanded to bipartite networks \citep{goyal2017inference}; it may be possible to apply similar approaches to extend the DCCM to include bipartite networks. Further work is also required to develop dynamic essential network properties whose functions do not depend only on features of the previous observed network, as making a Markov assumption can have significant impact on  epidemics models \citep{GWD12}. As shown in section 6, the proposed method provides greater flexibly than many existing network models in that it does not require the probability distribution of the dynamic essential network properties to conform to the Markov assumption.

\section*{Acknowledgments}

This research is supported by grants from the National Institutes of Health (R37 AI-51164). Conflict of Interest: None declared.

\pagebreak

\section{Appendix: Technical Details for DCCM}

The predicted networks were generated using a Metropolis-Hastings algorithm with target distribution based on equation (\ref{eq:prob_cc_networkmixing}). Use of Metropolis-Hastings algorithm requires evaluation of the acceptance probability, as described in equations (\ref{eq:prob_accept_DCCM}). Since (\ref{eq:eta_predict_3}) provides the probability mass function for $P_{\mathcal{C}_t}$, we only need to calculate $f(C_g, C_h)$. 

Though our analysis considers mixing based on only political party membership, the equations below are generalized to allow for mixing between individuals based on an arbitrary number of covariate patterns. We present the quantities for the four cases that must be evaluated in order to calculate $f(c_{\eta(g_{t}' \vert g_{t-1})}, c_{\eta(gp_{t} \vert g_{t-1})})$. Let edge $(i,j)$ be the required edge toggle to move from $g_t'$ to $gp_t$ and let $S^{l,k}(g) = \{E_{ij} : E_{ij} \in g, \DD{m}{i}{} = l, \mbox{ and } \DD{m}{j}{} = k\}$. The four cases are associated with whether $(i,j)$ exists in $g_t'$ or $g_{t-1}$ or both or neither. \vspace{.5cm}

\noindent \textit{Case 1:} $(i,j) \in g_t'$ and $(i,j) \in g_{t-1}$. Therefore,

\begin{equation} \label{eq:f_mixing_1a}
\eta_s^{l,k}(gp_t)*\DDg{M}{m_i}{}{g_t'} = \eta_s^{l,k}(g_t')*\DDg{M}{m_i}{}{g_t'} - I_{\{m_i = l, m_j = k\}},
\end{equation}

\noindent and

\begin{equation} \label{eq:f_mixing_1b}
\eta^{l,k}_d(gp_t \vert g_{t-1})*\DDg{M}{m_i}{}{g_t'} = \eta^{l,k}_d(g_t' \vert g_{t-1})*\DDg{M}{m_i}{}{g_t'} - I_{\{m_i = l, m_j = k\}}.
\end{equation}

\noindent Toggling any edge in $S^{l,k}(g_t') \bigcap S^{l,k}(g_{t-1})$ would satisfy equations (\ref{eq:f_mixing_1a}) and (\ref{eq:f_mixing_1b}); since this logic holds for any $g \in c_{\eta(g_{t}' \vert g_{t-1})}$ and $\vert S^{l,k}(g_t') \bigcap S^{l,k}(g_{t-1}) \vert$ is constant across $g \in c_{\eta(g_{t}' \vert g_{t-1})}$,

\begin{equation} \label{eq:f_function_mixing_1}
f(c_{\eta(g_{t}' \vert g_{t-1})}, c_{\eta(gp_{t} \vert g_{t-1})}) = \eta^{m_i,m_j}_d(g_t' \vert g_{t-1})*\DDg{M}{m_i}{}{g_t'}.
\end{equation}
\noindent \textit{Case 2:} $(i,j) \in g_t'$ and $(i,j) \notin g_{t-1}$. Therefore,

\begin{equation} \label{eq:f_mixing_2a}
\eta_s^{l,k}(gp_t)*\DDg{M}{m_i}{}{g_t'} = \eta_s^{l,k}(g_t')*\DDg{M}{m_i}{}{g_t'} - I_{\{m_i = l, m_j = k\}},
\end{equation}

\noindent and

\begin{equation} \label{eq:f_mixing_2b}
\eta^{l,k}_d(gp_t \vert g_{t-1})*\DDg{M}{m_i}{}{g_t'} = \eta^{l,k}_d(g_t' \vert g_{t-1})*\DDg{M}{m_i}{}{g_t'}. 
\end{equation}

\noindent Any edge from $S^{l,k}(g_t') / S^{l,k}(g_{t-1})$ can be toggled to satisfy equations (\ref{eq:f_mixing_2a}) and (\ref{eq:f_mixing_2b}). Again, because this reasoning holds for any $g \in c_{\eta(g_{t}' \vert g_{t-1})}$ and because $\vert S^{l,k}(g_t') / S^{l,k}(g_{t-1}) \vert$ is constant across $g \in c_{\eta(g_{t}' \vert g_{t-1})}$,

\begin{equation} \label{eq:f_function_mixing_2}
f(c_{\eta(g_{t}' \vert g_{t-1})}, c_{\eta(gp_{t} \vert g_{t-1})}) = \eta_s^{m_i,m_j}(g_t')*\DDg{M}{m_i}{}{g_t'} - \eta^{m_i,m_j}_d(g_t' \vert g_{t-1})*\DDg{M}{m_i}{}{g_t'}.
\end{equation}

\noindent \textit{Case 3:} $(i,j) \notin g_t'$ and $(i,j) \in g_{t-1}$. Therefore,

\begin{equation} \label{eq:f_mixing_3a}
\eta_s^{l,k}(gp_t)*\DDg{M}{m_i}{}{g_t'} = \eta_s^{l,k}(g_t')*\DDg{M}{m_i}{}{g_t'} + I_{\{m_i = l, m_j = k\}},
\end{equation}

\noindent and

\begin{equation} \label{eq:f_mixing_3b}
\eta^{l,k}_d(gp_t \vert g_{t-1})*\DDg{M}{m_i}{}{g_t'} = \eta^{l,k}_d(g_t' \vert g_{t-1})*\DDg{M}{m_i}{}{g_t'} + I_{\{m_i = l, m_j = k\}}.
\end{equation}

\noindent An edge from $S^{l,k}(g_{t-1}) / S^{l,k}(g_{t}')$ can be toggled to satisfy equations (\ref{eq:f_mixing_3a}) and (\ref{eq:f_mixing_3b}). Therefore,

\begin{equation} \label{eq:f_function_mixing_3}
f(c_{\eta(g_{t}' \vert g_{t-1})}, c_{\eta(gp_{t} \vert g_{t-1})}) = \eta_s^{m_i,m_j}(g_{t-1})*\DDg{M}{m_i}{}{g_t'} - \eta^{m_i,m_j}_d(g_t' \vert g_{t-1})*\DDg{M}{m_i}{}{g_t'}
\end{equation}

\noindent for similar reasons as the previous cases.

\noindent \textit{Case 4:} $(i,j) \notin g_t'$ and $(i,j) \notin g_{t-1}$. Therefore,

\begin{equation} \label{eq:f_mixing_4a}
\eta_s^{l,k}(gp_t)*\DDg{M}{m_i}{}{g_t'} = \eta_s^{l,k}(g_t')*\DDg{M}{m_i}{}{g_t'} + I_{\{m_i = l, m_j = k\}},
\end{equation}

\noindent and

\begin{equation} \label{eq:f_mixing_4b}
\eta^{l,k}_d(gp_t \vert g_{t-1})*\DDg{M}{m_i}{}{g_t'} = \eta^{l,k}_d(g_t' \vert g_{t-1})*\DDg{M}{m_i}{}{g_t'}.
\end{equation}

\noindent An edge from all possible edges connecting an $m_i$ node to an $m_j$ node that is not in $S^{l,k}(g_t') \bigcup S^{l,k}(g_{t-1})$ can be toggled to satisfy equations (\ref{eq:f_mixing_4a}) and (\ref{eq:f_mixing_4b}). Therefore,

\begin{equation} \label{eq:f_function_mixing_4}
f(c_{\eta(g_{t}' \vert g_{t-1})}, c_{\eta(gp_{t} \vert g_{t-1})}) = \DDg{M}{m_i, m_j}{}{g_t'}*\DDg{M}{m_i}{}{g_t'} - [\eta_s^{m_i,m_j}(g_{t-1})*\DDg{M}{m_i}{}{g_t'} - \eta^{m_i,m_j}_d(g_t' \vert g_{t-1})*\DDg{M}{m_i}{}{g_t'}],
\end{equation}

\noindent where

\begin{equation} \label{eq:f_function_mixing_4b}
\DDg{M}{m_i, m_j}{}{g_t'} = \left\{\begin{array}{lll}
[(\DDg{M}{m_i}{}{g_t'} * \DDg{M}{m_j}{}{g_t'}) - \eta_s^{m_i,m_j}(g_{t}')*\DDg{M}{m_i}{}{g_t'}] & if & m_i \neq m_j \\
{\DDg{M}{m_i}{}{g_t'} \choose 2} - \eta_s^{m_i,m_j}(g_{t}')*\DDg{M}{m_i}{}{g_t'} & if & m_i = m_j,
\end{array}\right.
\end{equation}

\noindent for similar reasons as the previous cases. The calculations for $f(c_{\eta(gp_{t} \vert g_{t-1})}, c_{\eta(g_{t}' \vert g_{t-1})})$ are similar to $f(c_{\eta(g_{t}' \vert g_{t-1})}, c_{\eta(gp_{t} \vert g_{t-1})})$.

\pagebreak

\bibliographystyle{nws}
\bibliography{paper2_bib}

\end{document}